\documentclass[12pt,a4paper]{iopart}
\usepackage{iopams,graphicx}
\usepackage{cite}
\newcommand{\la}{\langle}
\newcommand{\ra}{\rangle}

\newcommand{\diff}{{\rm d}}

\newcommand{\vp}{\boldsymbol{p}}

\newcommand{\vu}{\boldsymbol{u}}

\newcommand{\V}{ {\cal V} }

\newcommand{\Op}[1]{{\sf #1}} %Operator-Formatierung
\newcommand{\ox}{ \Op{x} } %Ort
\newcommand{\rhoo}[1]{\rhoo \left(#1 \right)}

\begin{document}
\title[Concept of an ionizing time-domain matter-wave interferometer ]{Concept of an ionizing time-domain  matter-wave interferometer}

\author{Stefan Nimmrichter$^1$, Philipp Haslinger$^1$,  Klaus Hornberger$^2$, and Markus Arndt$^1$}
\address{$^1$Vienna Center for Quantum Science and Technology (VCQ), University of Vienna, Faculty of Physics, Boltzmanngasse 5, 1090 Vienna, Austria}
\address{$^2$ Max-Planck Institute for the Physics of Complex Systems, N\"{o}thnitzer Str. 38, 01187 Dresden, Germany}
\ead{markus.arndt@univie.ac.at}

\begin{abstract}
We discuss the concept of an all-optical and ionizing matter-wave interferometer in the time domain. The proposed setup aims at testing the wave nature of highly massive clusters and molecules, and it will enable new precision experiments with a broad class of atoms, using the same laser system.
The propagating particles are illuminated by three pulses of a standing ultraviolet laser beam, which detaches an electron via efficient single photon-absorption.
Optical gratings may have periods as small as 80\,nm, leading to wide diffraction angles for cold atoms and to compact setups even for very massive clusters.
Accounting for the coherent and the incoherent parts of the particle-light interaction, we show that the combined effect of phase and amplitude modulation of the matter waves gives rise to a Talbot-Lau-like interference effect with a characteristic dependence on the pulse delay time.
\end{abstract}

%42.50.-p, %Quantum optics
%03.75.-b  %Matter waves
%33.80.-b, %Photon interactions with molecules (see also 42.50 Quantum optics)
%03.65.Ta  %Foundations of quantum mechanics; measurement theory
%03.75.Dg  %Atom and neutron interferometry in quantum
%39.20.+q  % Atom interferometry techniques (see also 03.75.-b Matter waves, and mechanics)
%03.65.-w  % Quantum mechanics
%\vspace{2pc}
%\submitto{\NJP}
\maketitle

\tableofcontents
\title[Matter wave interferometry with photoionization gratings in the time-domain]{}

\section{Introduction} \label{sec:intro}
\subsection{Interferometry with molecules}
Over the last decades, matter-wave interferometry has developed into a highly active research field,
ranging from the foundations of physics to quantum enhanced precision measurements.
Beam splitters, mirrors, diffraction gratings, traps or wave guides are nowadays readily available for electrons~\cite{Hasselbach2010}, neutrons \cite{Rauch2000} and atoms~\cite{Cronin2009}, but they  still represent a substantial challenge for large clusters or molecules.

Diffraction and interference experiments with molecules were started   systematically only in the mid-1990s and led to the exploration of far-field diffraction effects at nanomechanical~\cite{Schöllkopf1994,Arndt1999,Bruhl2002} and optical gratings~\cite{Nairz2001}, as well as at opaque discs~\cite{Reisinger2009}. Interferometers were operated using nanogratings in a Mach-Zehnder configuration~\cite{Chapman1995}, running laser waves in a Ramsey-Bord{\'e} arrangement~\cite{Borde1994,Lisdat2000},  mechanical gratings in a near-field Talbot-Lau design~\cite{Brezger2002,Hackermuller2003} or, more recently, in a combination of mechanical and optical phase gratings~\cite{Gerlich2007,Hornberger2009}.

All these designs have their merits and drawbacks. While far-field experiments are conceptually appealing and allow one to spatially separate the diffraction orders, it was already pointed out by Clauser \cite{Clauser1997a} that the use of a near-field configuration, such as the Talbot-Lau interferometer, offers several advantages, in particular much reduced collimation and coherence requirements and therefore increased signals.

\subsection{Optical elements for matter wave interferometry}
Nanomechanical gratings, which serve to block a part of the particle beam, are the most natural diffraction elements for matter waves. Using modern lithographic techniques, it is nowadays possible to nanofabricate e.g.~silicon nitride structures with a precision that guarantees a predefined slit period to within a few Angstroms, even over millimeter sized areas~\cite{Savas1995}. Such masks are of great importance for many applications in atom~\cite{Keith1988,Carnal1991} and electron interferometry \cite{Gronniger2005,Cronin2006} since they do not rely on any internal particle property.

However, for highly polarizable and slow particles,
the presence of dispersion forces near the grating walls
becomes increasingly important~\cite{Grisenti1999,Brezger2002,Bruhl2002,Perreault2006}.
These van der Waals or Casimir-Polder interactions introduce a phase shift with a strong position and velocity dependence \cite{Hornberger2004,Nimmrichter2008b}.
The force may even remove a substantial part of the molecules from the beam when they get too close to the surface.

In contrast to that, optical standing-wave gratings can neither be destroyed nor clogged by
large particles. 
Narrow-band lasers allow one to define the grating period
with high precision
and the grating transmission function, defined by the particle-light interaction,
can be controlled  and modulated in situ and on a short time scale via the laser intensity~\cite{Steane1995,Cahn1997}.

\subsection{From phase gratings to absorptive gratings of light}
The use of optical phase gratings is by now well established both in atom interferometry (e.g.~\cite{Martin1988,Pfau1994,Szriftgiser1996a,Cahn1997,Keller1997}), with electron beams \cite{Freimund2001}, and also for complex molecules~\cite{Nairz2001,Gerlich2007}.
Interferometry with phase gratings requires a prior effort in preparing sufficient transverse coherence in the particle beam.
This is naturally provided by an atomic Bose-Einstein condensate\cite{Deng1999} or if the diffracting particles start already  in a  periodic arrangement predefined by a potential~\cite{Turlapov2005}.
A spatially extended and incoherent molecular beam requires, however, to use an absorptive grating prior to the diffracting phase mask to prepare some transverse coherence.
An elegant way of realizing
an absorptive optical grating for atoms is to pump them into a dark state, i.e.~into an undetected internal
state. This strategy works rather well for atoms~\cite{Abfalterer1997,Cohen2000}, but clusters and molecules usually do not offer the needed addressability of selected states.

In our present work we focus on the implementation of absorptive gratings via single-photon ionization  in the anti-nodes of an ultraviolet standing light wave \cite{Reiger2006}. Modern (V)UV lasers readily provide the photon energy required to surpass the ionization threshold of many kinds of neutral atoms, clusters and molecules.
The optical grating will imprint a nanostructured periodic amplitude profile onto the particle beam, very much like a material mask and separate charged and neutral particles.
We study, in particular, the sequence of three ionizing optical grating pulses
in the time domain.
The delay between the pulses serves as a control parameter, while the spatial positions of the gratings are fixed.

The high time resolution of modern lasers allows us to take advantage of all the benefits
that are shared by other
atom interferometers in the time domain~\cite{Kasevich1991a,Szriftgiser1996a, Cahn1997}: the interference fringe positions  can be largely independent of the particle velocity and
the interference pattern can be scanned with very high accuracy and without any moving mechanical parts.
While the idea of ionization gratings is here worked out for massive clusters,
the method is naturally extended to atoms, as well.

\section{Optical time-domain ionizing matter interferometer} \label{sec:setup}

\begin{figure}
\includegraphics[width=\columnwidth]{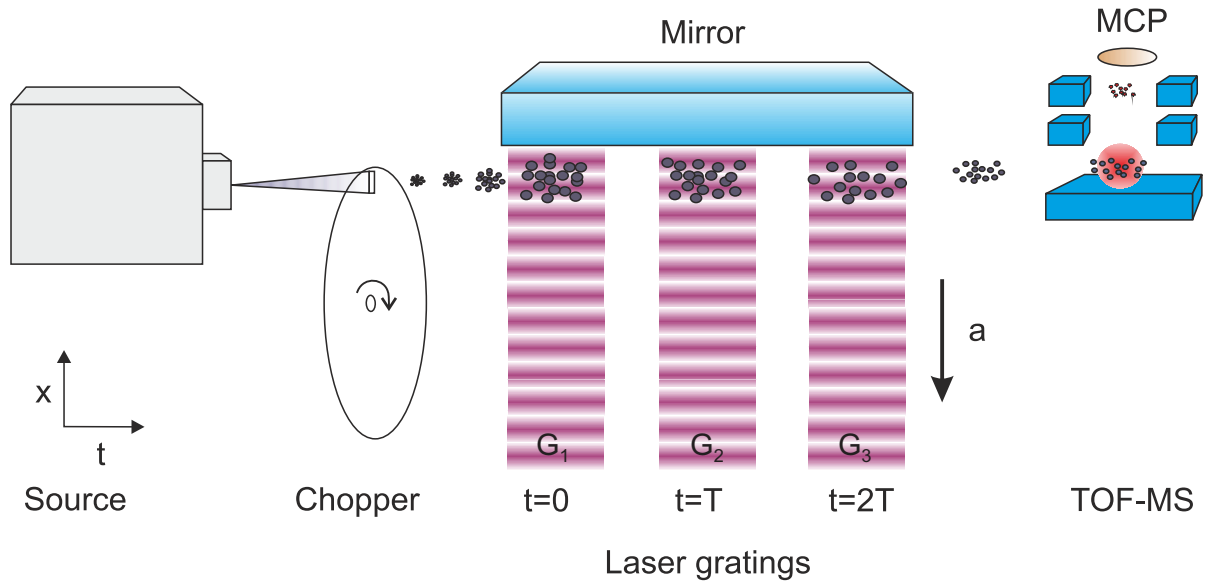}
\caption{\label{fig:pulsescheme}
Experimental layout and pulse sequence for an optical time-domain ionizing matter-wave (OTIMA) interferometer. A beam of nanoparticles
is chopped into bunches, each of which is subsequently illuminated by the same three laser pulses (G$_1$, G$_2$ and G$_3$), separated by a variable but equal delay time $T$ on the order of the Talbot time $T_{\rm T}$.
The retro-reflection of the laser light at a common plane mirror generates a phase-stable standing wave. 
The laser wave length is sufficiently short
to ionize the particles with high probability at the antinodes after absorbtion of a single photon. 
The ions are removed by a homogeneous electric field.
The remaining neutral particles fly into a time-of-flight mass spectrometer (TOF-MS), where they are photo-ionized, accelerated and detected by a multi-channel plate (MCP). This scheme allows one to post-select the clusters with a mass selectivity of better than 0.1\,\% and to perform the experiment with a large number of different masses simultaneously. A quantum interference fringe pattern is observed in the count rate by varying the pulse separation $T$, provided the clusters are exposed to an additional constant acceleration $a$, e.g.~due to gravitation or a constant electrical field gradient.}
\end{figure}

In the following, we will consider particles that are much smaller than the laser wave length, and refer to them as `clusters', even though the concept also applies to atoms or some molecules.
The particle-light interaction is then governed by two parameters, the (real) optical dipole polarizability $\alpha$ and the absorption cross section $\sigma_{\rm abs}$ at the laser frequency.

\subsection{Interferometric setup}
\label{sec:basics}

The layout and time sequence of an Optical Time-domain Ionizing Matter-wave (OTIMA) interferometer are depicted in Fig.~\ref{fig:pulsescheme}. A pulsed cluster beam passes alongside a plane mirror which serves to retro-reflect the beam of an ultraviolet (UV) laser. The light source illuminates each cluster cloud in  three short pulses, separated by the equal time delay $T$. We denote them by G$_1$, G$_2$ and G$_3$.
The laser beam waist is chosen such that all clusters are covered equally by the standing light waves, thus ensuring that all particles with the same distance to the mirror experience the same light intensity. Depending on the velocity of the cluster beam, this can be realized by a single laser of high repetition rate or three pulses that follow the particle cloud along its direction of motion.

The particles are also subjected to a (weak) constant force perpendicular to the mirror plane, e.g.~gravity or an electric field with $\nabla E^2=$const.
The field will also serve to remove all ions from the beam. 
Since all relevant forces only depend on the cluster-mirror distance, we restrict our attention to this $x$-axis.

A grating pulse acts in several ways: First, the optical dipole potential imprints a sinusoidal phase modulation  onto the center-of-mass state of the clusters.
The period $d$ is given by half the laser wavelength, $d=\lambda_{\rm L}/2$.  The maximal phase shift $\phi_0$ is proportional to the real part of the optical polarizability $\alpha$ and to the laser pulse energy.
Second, even a single absorbed photon will ionize a given cluster. The
probability of such ionization events is again an oscillating
function of the transverse position $x$. The average number $n_0$ of absorbed photons in the antinodes   is proportional to the absorption cross section $\sigma_{\rm abs}$ and to the laser pulse energy. The spatially periodic ionization of the clusters, and their removal by the electric field, then results in an amplitude modulation of the particle beam, much like in a material grating.
It is important that a cluster does indeed ionize with high probability after the absorption of a single photon.
Photoabsorption results in a momentum kick of $\pm h/\lambda_{\rm L}$ with a random sign.  This could  blur the interference pattern if the particles were not ionized and removed. 
A third effect is the radiation pressure exerted on the clusters due to Rayleigh scattering.
This is a potential source of decoherence since the scattered photons impart momentum kicks onto the clusters which are comparable to the `grating momentum' $\Delta p_G= h/d$  but randomly directed (see Sect.~\ref{sec:scattering}). 
Behind the interferometer the remaining neutral clusters are detected mass-selectively. The
interference pattern will manifest itself in a characteristic dependence of the count rate on the particle mass $m$, the pulse separation $T$, the external acceleration $a$ and also the laser pulse intensity.

\subsection{Near field interference}
Near-field Talbot-Lau interferometers (TLI) are particularly well suited for exploring quantum wave mechanics with massive particles and short de Broglie wave lengths \cite{Clauser1997a,Brezger2002,Hornberger2004,Gerlich2007,Nimmrichter2008b}.
In our time-domain version, the laser pulses modify the $x$-component of the motional cluster state in 
the same way as the three gratings of a TLI 'in space'.
The two concepts are  related to each other by a change of the reference frame.

Compared to a mechanical Talbot-Lau setup, the optical analog adds substantial control due to the precisely defined delay between the pulses and the possibility of tuning the pulse strengths individually. This allows one to overcome source imperfections, in particular the longitudinal velocity spread in the beam, and to optimize the fringe visibility.
In contrast to the spatial near-field interferometer, which is  characterized by the Talbot distance
 $L_{\rm T} =  d^2/\lambda$, the characteristic scale in the time domain is given by the \textit{Talbot time}  \cite{Cahn1997}
\begin{equation}
\label{eqn:TL}
T_{\rm T} = m d^2/h
\end{equation}
This is the typical time for the self-imaging of a pulsed diffraction grating under coherent illumination.

In practice, the $x$-momentum distribution $\Delta p_x$ of the particle cloud will be much broader than the `grating momentum' $h/d$. The role of G$_{1}$ at $t=0$ is therefore reduced to acting as a mask that prepares spatial coherence by only transmitting particles from many parallel but very narrow 'sources' in the nodes of the standing laser wave.
Some transverse coherence thus emerges after time $t=T$, when the second grating pulse G$_{2}$ is applied. The amplitude and phase modulation at G$_{2}$ then leads to a resonant spatial modulation of the cluster density at time $t=2T$ \cite{Cahn1997,Dubetsky1999}.
The normalized particle density (Sect.~\ref{sec:theory}) then takes the form
\begin{equation}
\label{eqn:w3TL}
 w_{2T}(x) = \frac{1}{d} \sum_{\ell=-\infty}^\infty  B_{-\ell}^{(1)} (0)
B_{2\ell}^{(2)} \left( \ell \frac{T}{T_{\rm T}} \right)
\exp \left( \frac{2\pi \rmi \ell (x-a T^2)}{d} \right) .
\end{equation}
This pattern is again periodic in $x$, with period $d$ but it is shifted in the presence of an acceleration $a$.

The Fourier components characterizing the fringe pattern  (\ref{eqn:w3TL}) can be specified in terms of the
\textit{Talbot-Lau coefficients} of the $k$-th grating \cite{Hornberger2009},
\begin{equation}
 B_{m}^{(k)} (\xi) = e^{-n_0^{(k)} / 2} \left[ \frac{\zeta_{\rm coh} - \zeta_{\rm ion}}{\zeta_{\rm coh} + \zeta_{\rm ion}} \right]^{m/2} J_m \left( {\rm sgn} (\zeta_{\rm coh} + \zeta_{\rm ion}) \sqrt{\zeta_{\rm coh}^2 - \zeta_{\rm ion}^2} \right), \label{eqn:TLcoeffgeneral}
\end{equation}
involving the expressions
\begin{eqnarray}
\label{eqn:zeta-abs}
\zeta_{\rm ion} &=& n_0^{(k)} \cos (\pi \xi) / 2,
\\
\label{eqn:zeta-phase}
\zeta_{\rm coh} &=& \phi_0^{(k)} \sin (\pi \xi).
\end{eqnarray}
Here  $J_m$ is the Bessel function of the first kind and $n_0^{(k)}$ and $\phi_0^{(k)}$ are the above mentioned grating pulse parameters, specifying the maximal mean number of absorbed photons and the maximum phase shift, respectively. They are defined in Sect.~\ref{sec:theory} by Eqs.~(\ref{eqn:I}) and (\ref{eqn:beta}).

Note in Eq.~(\ref{eqn:w3TL}) how the dependence on $\xi = \ell T / T_{\rm T}$ in the second grating coefficient modulates the parameters (\ref{eqn:zeta-abs}) and (\ref{eqn:zeta-phase})  of ionization and phase shift, and thus relates the pulse timing $T$ to the Talbot time $T_{\rm T}$. It is in this functional dependence that the quantum scale $T_{\rm T}$ enters, leading to the rich fringe structures characteristic of near field interference phenomena.

The components attributed to the first grating mask ($k=1$) take the simpler form
\begin{equation}
 B_{m}^{(k)} (0) = \exp\left( -\frac{n_0^{(k)}}{2} \right)  I_m \left( -
\frac{n_0^{(k)}}{2} \right), \label{eqn:ATL}
\end{equation}
where $I_m$ is the modified
Bessel function of the first kind.  
Note that this factor depends only on the photon absorption $n_0^{(k)}$, not on the phase shift. In the limit of small photon absorption $n_0^{(1)} \to 0$, we find $B_m^{(1)}(0) = \delta_{m,0}$. This implies that if the first optical grating does not ionize the clusters the final density distribution (\ref{eqn:w3TL}) will be as uniform as the initial particle cloud, even if there is a substantial additional phase shift.
In the case of equal time delays $T$ between the pulses, the first grating must therefore be ionizing, i.e absorptive,  in order to generate a fringe pattern.

Near-field interference may arise behind pure phase gratings, under some circumstances:
As discussed in Sec.~\ref{sec:theory}, coherent rephasing effects can be expected at times $T+\tau$ after the second pulse, with  $|\tau| \ll T$.
Such transient near-field diffraction phenomena were observed in several time-domain experiments~\cite{Cahn1997,Wu2005a,Tonyushkin2006a} where the Bragg reflection of the third laser pulse was used to detect the reconstructed atomic density pattern.
In contrast to that, our present proposal
uses the symmetric  Talbot-Lau recurrence with its high visibility and robustness,
as required for slow cluster beams detected by a third absorptive mask.

\subsection{Fringe visibility} \label{sec:detection}
The fringe pattern (\ref{eqn:w3TL}) is finally probed by the third laser field  G$_{3}$
which ionizes and extracts the clusters in its antinodes and transmits
the remaining neutrals to the detector. Their fraction is proportional to
\begin{equation}
 S = \sum_{\ell=-\infty}^\infty  \underbrace{B_{-\ell}^{(1)} (0)
B_{2\ell}^{(2)} \left( \ell \frac{T}{T_{\rm T}} \right) B_{-\ell}^{(3)} (0)}_{\equiv S_{\ell}}\exp \left( \frac{2\pi \rmi \ell a T^2}{d} \right). \label{eqn:sigTL}
\end{equation}
When recorded as a
function of the transverse acceleration $a$, this periodic signal has a form similar to the density pattern (\ref{eqn:w3TL}). The Fourier coefficients now merely contain in addition the components $B_{n}^{(3)}(0)$ of the third grating pulse. A useful interference pattern can be recorded even if the force is constant
as in the presence of gravity $a=g$: 
In this case one varies the pulse delay $T$ on the scale $\Delta T \lesssim d / 2gT$. If this is much smaller than the Talbot time $T_{\rm T}$ the resulting pattern is practically periodic in $T^2$ since the Fourier coefficients then hardly change on that scale. This is naturally the case for massive clusters, while one requires a high multiple $M \gg 1$ of the Talbot time, $T = MT_{\rm T}$, for atoms and light molecules.

The fringe visibility of the periodic signal (\ref{eqn:sigTL}) is conventionally defined as
\begin{equation}
\V = \frac{S_{\max} - S_{\min}}{S_{\max} + S_{\min}} \label{eqn:vis}.
\end{equation}
In practice, the pattern is often close to sinusoidal, allowing one to describe the experimental observation by fitting a $d$-periodic sine curve with offset $S_0$ and amplitude $2S_1$ to the measured data. It is then more appropriate to consider the
`sinusoidal visibility' $\V_{\sin}$, the visibility associated to the first Fourier component.
For the expected signal (\ref{eqn:sigTL}), and after the insertion of  (\ref{eqn:TLcoeffgeneral}) and (\ref{eqn:ATL}), it is given by
\begin{equation}
\fl \V_{\sin} =  \frac{2|S_1|}{S_0} = 2 \frac{I_1 \left( n_0^{(1)} / 2 \right)}{I_0 \left(
n_0^{(1)} / 2 \right)}
\cdot
 \frac{\left|(\zeta_{\rm coh} - \zeta_{\rm ion}) J_2 \left( \sqrt{\zeta_{\rm
coh}^2 - \zeta_{\rm ion}^2} \right)\right|}{|\zeta_{\rm coh} + \zeta_{\rm ion}|\, I_0 \left( n_0^{(2)} / 2 \right)}
\cdot
\frac{I_1 \left( n_0^{(3)} / 2 \right)}{I_0 \left( n_0^{(3)} / 2 \right)}. \label{eqn:vissinTL2}
\end{equation}
Note that the sinusoidal visibility may assume values larger than one, while the conventional visibility (\ref{eqn:vis}) by definition cannot exceed $100\%$.

\subsection{Quantum or classical patterns?}
It is important to keep in mind that the observation of fringes in a two-grating or three-grating setup  is, by itself,  not yet a conclusive proof of quantum interference, since moir\'{e}-type shadow patterns can also be created by particles moving on classical trajectories. The genuine quantum origin of the patterns is revealed by their characteristic dependence on the quantum scale (\ref{eqn:TL})
and the detailed functional dependence of the fringe visibility on various parameters.
 It is encoded, formally, in the oscillating $\xi$-dependence of the Talbot-Lau coefficients (\ref{eqn:TLcoeffgeneral}) for the second grating pulse, while the actions of G$_{1}$ and G$_{3}$
 are the same  in the quantum and the classical formulation. Calculating the deflection of a classical trajectory under the influence of a standing light-wave one is led to classical analogs of the Talbot-Lau coefficients~\cite{Hornberger2009}. Their form is identical to the quantum coefficients (\ref{eqn:TLcoeffgeneral}), except for the definitions (\ref{eqn:zeta-abs}) and (\ref{eqn:zeta-phase}) of the parameters $\zeta_{\rm ion}$ and $\zeta_{\rm coh}$. They are
now simply given by $\zeta_{\rm ion} = n_0^{(k)}/2$ and $\zeta_{\rm coh} = \phi_0^{(k)} \pi \xi$; in particular, they do not oscillate with $\xi$.

In agreement with intuitive expectations, the classical and the quantum description converge in the limit of large masses, $m\to \infty$, or for short pulse separations, $T \to 0$, where the argument of the Talbot-Lau coefficients (\ref{eqn:TLcoeffgeneral}) tends to zero. They do differ markedly for finite masses and  times. Genuine quantum interference becomes clearly distinguishable from any classical dynamics when the pulse delay time $T$ exceeds at least the Talbot time, $T \gtrsim T_{\rm T}$.

\subsection{Expected visibilities for high-mass clusters} \label{sec:numerics}

We now discuss how the expected interference visibility (\ref{eqn:vissinTL2}) depends on the laser pulse parameters and on the optical cluster properties for high-mass metal clusters. The photon absorption parameters $n_0^{(k)}$ grow linearly with the laser intensities; they  determine the 'survival factors' $B_0^{(k)}(0)$ for each grating pulse, i.e.~the spatially averaged probabilities of a particle not to be ionized during the $k$-th pulse.
The optimal laser power is therefore a compromise: Large photon numbers may increase the interference visibility, but they also decrease the number of clusters that arrive at the detector, $S_0 = B_0^{(1)}(0) B_0^{(2)}(0) B_0^{(3)}(0)$.

While the first and the third grating pulse are fully characterized by the absorption parameters $n_0^{(1)}$ and $n_0^{(3)}$, the second pulse shapes the interfering matter wave in amplitude and phase with parameters $n_0^{(2)}$ and $\phi_0^{(2)}$.
To account for the specific optical properties of the interfering particle independently of the laser intensity it is therefore convenient to use the dimensionless ratio 
\begin{equation}
 \beta = \frac{n^{(2)}_0}{2\phi_0^{(2)}} = \frac{\lambda_{\rm L} \sigma_{\rm abs}}{8\pi^2 \alpha}
\end{equation} 
with $\lambda_{\rm L}$ the laser wave length. We assume this to be 157\,nm, as will be justified below.

\begin{figure}
\includegraphics[width=\columnwidth]{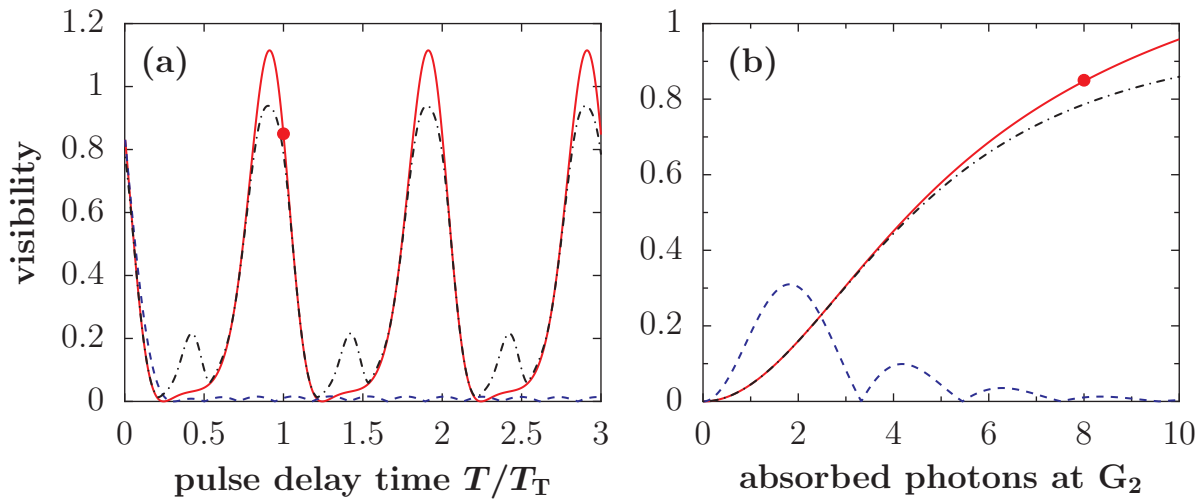}
\caption{\label{fig:visqmklass} (a) Simulation of the `sinusoidal' quantum interference visibility $\V_{\sin}$ (solid line), the conventional quantum contrast $\V$ (dash-dotted line), and the classical expectation for the fringe contrast (dashed line) as a function of $T/T_{\rm T}$ for  $n_0^{(2)}=8$. (b) The same contrast curves as a function of $n_0^{(2)}$ for fixed $T = T_{\rm T}$. The simulation assumes the material characteristics of gold clusters, $\beta = 1.0$.}
\end{figure}

In order to visualize the expected interference patterns, we keep the first and third grating intensities fixed at $n_0^{(1)}=n_0^{(3)}=8$ and choose the cluster parameter $\beta=1.0$ to be characteristic for large gold clusters, essentially independently of the particle size.
Figure \ref{fig:visqmklass}(a) depicts the sinusoidal visibility as a function of the delay time $T$ in units of the Talbot time $T_{\rm T}$ for $n_0^{(2)}=8$. Panel (b), on the other hand, shows the fringe contrast at $T=T_{\rm T}$ when the intensity of G$_2$ is varied between  $n_0^{(2)}=0$ and  $n_0^{(2)}=10$.
Both the quantum prediction (solid and dash-dotted curves) as well as the classical prediction (dashed curves) are included, and the circles in (a) and (b) identify the point of equal time and power in both graphs. We observe pronounced visibility peaks recurring at multiples of the Talbot time, a feature clearly absent in the classical calculation. In the following we therefore consider only the quantum case. Note also that the sinusoidal visibility is not always a sufficient approximation to the full fringe contrast (\ref{eqn:vis}). As seen in Fig.~\ref{fig:visqmklass}, it can overestimate the proper fringe amplitude in high-contrast regions (solid line), but it can also vanish where the total contrast is still finite (dash-dotted line). Depending on the experimental data analysis, higher-order Fourier coefficients may need to be accounted for in such cases.

\begin{figure}
\includegraphics[width=12cm]{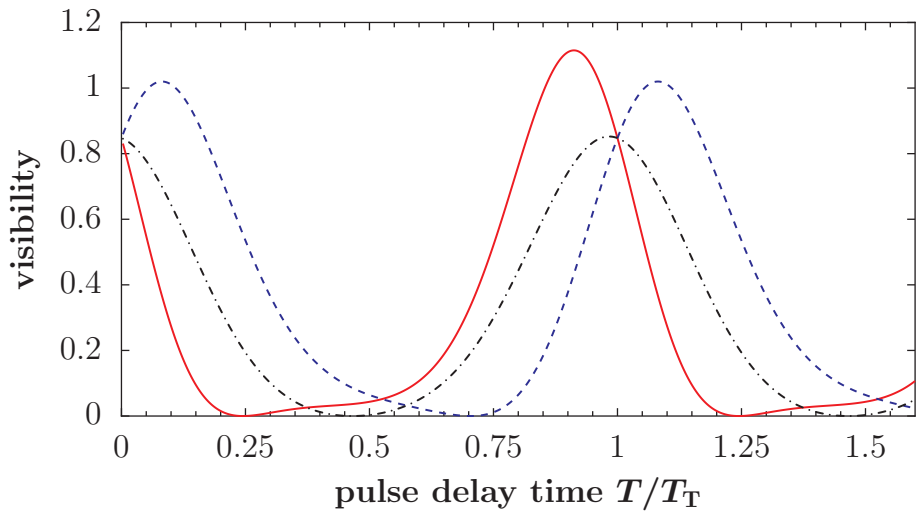}
\caption{\label{fig:visbeta} Predicted quantum interference contrast as a function of the pulse delay time $T/T_{\rm T}$ for different cluster materials, i.e.~for different values of the parameter $\beta$. The solid line corresponds to gold ($\beta=1.0$), the dashed line to cesium ($\beta=-1.3$) and the dashed-dotted line to silver ($\beta=9.2$). The $\beta$-values are obtained from Eq.~(\ref{eqn:beta}), using dielectric functions of the bulk materials \cite{Palik1998}. The power of the second pulse is set to $n_0^{(2)}=8$.
}
\end{figure}

Figure \ref{fig:visbeta} illustrates the influence of the material parameter $\beta$ on the sinusoidal visibility as a function of the pulse separation for three different materials, cesium, gold and silver clusters. Their $\beta$-values are representative for most ionizable cluster materials. The peaks of the visibility curves resize and move away from integer multiples of the Talbot time depending on the magnitude and sign of $\beta$. The general shape, however, is mainly determined by the number of absorbed photons $n_0^{(2)}$ and stays roughly the same for all $|\beta| \gtrsim 1$.
The negative value $\beta=-1.3$ for cesium implies that it is a low field seeker at the wavelength $\lambda_L=157\,$nm.

\begin{figure}
\includegraphics[width=\columnwidth]{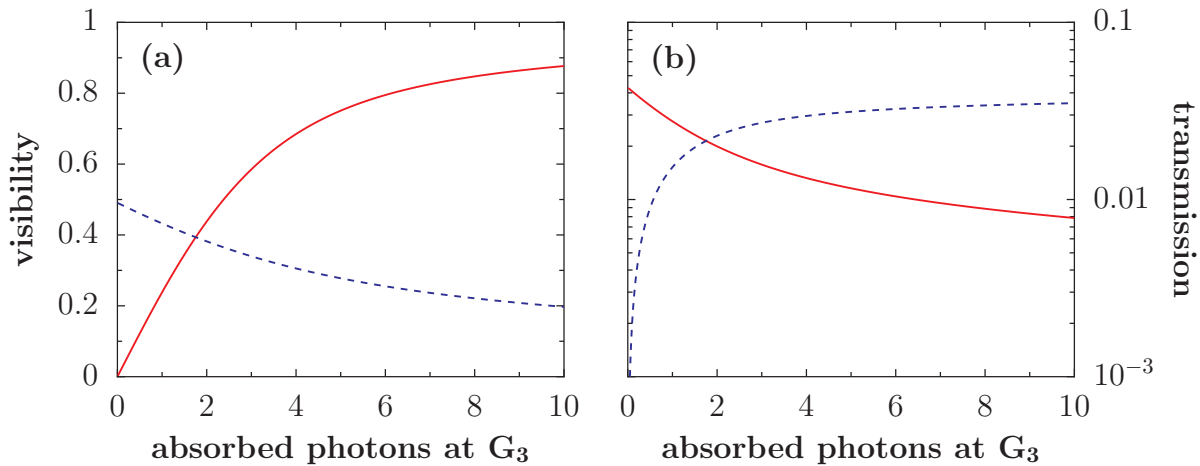}
\caption{\label{fig:grating13} (a) Computed quantum interference contrast and (b) transmissivity when the third grating laser power is varied. The power is expressed in terms of the number of absorbed photons. For the solid line in both plots the final signal is given by the transmitted neutral particles. The dashed line represents the case of an `inverted' grating  G$_3$. Here, the ions are counted directly. The increasing laser power then broadens the effective slit width of the third grating pulse.
}
\end{figure}

Finally, to see the effects of large laser powers we plot in Figure~\ref{fig:grating13}(a) the quantum contrast and in (b) the transmissivity as a function of the light intensity in G$_3$, while keeping the other parameters at the values of the marked points of Fig.~\ref{fig:visqmklass}. This case corresponds to the solid lines in Fig.~\ref{fig:grating13}. While the visibility grows with increasing laser power, the transmissivity decreases; the same holds for an intensity variation of G$_1$. In this regard, it is notable that the experiment can also be set up such that the count rate increases with increasing laser power in the third pulse, by counting the charged clusters directly. Even though this `inverse grating' configuration optimizes the detected cluster flux, it again reduces the fringe contrast, as shown by the dashed lines in Figure~\ref{fig:grating13}, since now any increase in laser intensity will broaden the `effective slit width' in G$_3$.

\section{Theoretical description} \label{sec:theory}

In the following section we recapitulate the description of complex gratings that are both ionizing and phase shifting. They are combined to the three-grating interferometer which we then model in phase space~\cite{Wigner1932,OzoriodeAlmeida1998,Schleich2001,Hornberger2004,Nimmrichter2008} to reconstruct the final particle distribution and the detected interference signal.
Finally, we also incorporate elastic laser light scattering.

\subsection{Complex light gratings} \label{sec:gratings}
Given a normalized transverse mode profile $f(y,z)$, with $\int \diff y \diff z \, f(y,z)=1$, of the laser beam the three-dimensional intensity distribution of the standing light wave is described by
\begin{equation}
 I\left( x,y,z,t \right) = 4 P_{\rm L} (t) f(y,z) \cos^2 \left( \frac{2 \pi x}{\lambda_{\rm L}} \right) , \label{eqn:I}
\end{equation}
where the total absorption is governed by the pulse energy $E_{\rm L} = \int \diff t \, P_{\rm L} (t)$. There is no need to average over the spatial laser beam profile since we assume the clusters to be centered in the laser beam. In the case of a Gaussian laser beam the profile function is described by
\begin{equation}
 f(y,z) = \frac{2}{\pi w_y w_z} \exp \left(-\frac{2y^2}{w_y^2} - \frac{2z^2}{w_z^2} \right).
\end{equation}
The cloud extension is of even lesser importance for pulsed lasers prepared with a flat-top spatial intensity profile.

If the particles are effectively at rest during the nanosecond pulse duration, we can assign a complex \textit{grating transmission function} $t^{(k)}(x)$ to each laser pulse $k=1,2,3$, which transforms a particle's wave function $\psi(x) \to t^{(k)}(x) \psi(x)$, according to a scattering model~\cite{Nimmrichter2008}. The modulus square $|t^{(k)}(x)|^2 \leq 1$ then gives the probability for a particle to be transmitted by the grating.
The mean number of absorption processes per cluster and pulse reads
\begin{equation}
 n^{(k)} (x) = \frac{4 \sigma_{\rm abs} E_{\rm L}^{(k)} \lambda_{\rm L}}{h c} f(0,0) \cos^2 \left( \frac{\pi x}{d} \right) \equiv n_0^{(k)} \cos^2 \left( \frac{\pi x}{d} \right), \label{eqn:n0}
\end{equation}
with the ionization cross section $\sigma_{\rm abs}$. The pulse energy $E_{\rm L}^{(k)}$ can be varied from pulse to pulse. Given the dielectric function $\varepsilon = \varepsilon_1 + \rmi \varepsilon_2$ of the cluster material the absorption cross-section of a spherical sub-wavelength cluster reads \cite{Kreibig1995}
\begin{equation}
 \sigma_{\rm abs} = 4 \pi R^3 \frac{2\pi }{\lambda_{\rm L}} \, {\rm Im} \left( \frac{\varepsilon - 1}{\varepsilon + 2} \right) = \frac{18\pi m }{\varrho \lambda_{\rm L}} \frac{\varepsilon_2}{(\varepsilon_1+2)^2 + \varepsilon_2^2},
 \end{equation}
where $R$ is the cluster radius and $\varrho = 3m / 4\pi R^3$ its mass density.

The probability for not absorbing any photon during the light pulse is $|t^{(k)}(x)|^2 = \exp (-n^{(k)}(x))$. This can be expanded in a Fourier series
\begin{equation}
 |t^{(k)}(x)|^2 = \sum_{n=-\infty}^\infty B_n^{(k)}(0) \exp \left( \frac{2\pi \rmi n x}{d} \right), \label{eqn:A}
\end{equation}
where the Fourier coefficients are given by Eq.~(\ref{eqn:ATL}).

In addition, the transmission function $t^{(k)}(x)$ also carries a phase due to the dipole energy \cite{Brezger2003,Hornberger2004,Nimmrichter2008}. It is mediated by the real part of the particle's dipole polarizability $\alpha_{\rm SI} (\lambda_{\rm L})$ at the laser wavelength, which we
use in volume units,
 $\alpha = \alpha_{\rm SI} (\lambda_{\rm L}) / 4\pi \varepsilon_0$. For sub-wavelength clusters it is given by
\begin{equation}
 \alpha = R^3 \, {\rm Re} \left( \frac{\varepsilon - 1}{\varepsilon + 2} \right) = \frac{3 m}{4\pi \varrho} \frac{\varepsilon_1^2 + \varepsilon_2^2 + \varepsilon_1 - 2}{(\varepsilon_1+2)^2 + \varepsilon_2^2}.
\end{equation}
We obtain the phase shift  $\phi^{(k)}(x) = \phi_0^{(k)} \cos^2 (\pi x /d)$, with $\phi_0^{(k)} = n_0^{(k)} / 2\beta$ and %
\begin{equation}
\label{eqn:beta}
 \beta = \frac{\lambda_{\rm L} \sigma_{\rm abs}}{8\pi^2 \alpha} = \frac{3 \varepsilon_2}{\varepsilon_1^2 + \varepsilon_2^2 + \varepsilon_1 - 2}.
\end{equation}
This leads to the total grating transmission function
\begin{equation}
 \fl t^{(k)}(x) = \exp \left( - \frac{n^{(k)}(x)}{2} + \rmi \phi^{(k)}(x) \right) = \exp \left[ \left(- \frac{n^{(k)}_0}{2} + \rmi \phi^{(k)}_0  \right) \cos^2 \left( \frac{\pi x}{d} \right) \right]. \label{eqn:t}
\end{equation}
Its Fourier coefficients are
\begin{equation}
 b_n^{(k)} = \exp \left( - \frac{n^{(k)}_0}{4} + \rmi \frac{\phi^{(k)}_0}{2} \right) I_n \left( - \frac{n^{(k)}_0}{4} + \rmi \frac{\phi^{(k)}_0}{2} \right). \label{eqn:bn}
\end{equation}
In contrast to the Kapitza-Dirac Talbot-Lau interferometer~\cite{Gerlich2007,Hornberger2009} where photoabsorption led to a contrast-reducing momentum recoil instead of the removal of molecules from the beam, both the phase-imprint and the single-photon ionization in G$_{2}$ are independently capable of causing matter-wave interference in our present OTIMA interferometer.

\subsection{Phase space model} \label{sec:phasespacemodel}
Starting from an initially bunched ensemble of particles at $t=0$ we are now going to derive an expression for its final state at $t=T_1+T_2$ using a one-dimensional phase space description of its time evolution~\cite{Hornberger2004,Nimmrichter2008,Hornberger2009}. The motional density matrix $\rho$ can be rewritten as the Wigner function
\begin{equation}
 w(x,p) = \frac{1}{2\pi \hbar} \int \diff s \, \exp \left( \rmi p s/\hbar \right)  \la x - \mbox{$\frac{1}{2}$} s |\rho |x + \mbox{$\frac{1}{2}$} s \ra, \label{eqn:wigner}
\end{equation}
which is a real function of the phase space coordinates $x$ and $p$. We assume the initial Wigner function to be spatially constant on the scale of the grating period $d$, since the incident cluster cloud with spatial extension $\Delta x$ is uniformly distributed over many periods of the standing wave, $\Delta x \gg d$. The momentum dependence is given by the one-dimensional marginal distribution
\begin{equation}
D(p) = \int \diff p_y \diff p_z \, \mu (p,p_y,p_z), \label{eqn:D}
\end{equation}
where $\mu (\vp)$ represents the three-dimensional momentum distribution of the particle cloud. The Fourier transform
\begin{equation}
 \widetilde{D} (s) = \int \diff p \, D(p) \exp \left(- \rmi p s/\hbar \right) \label{eqn:DFT},
\end{equation}
which is normalized to $\widetilde{D} (0) = \int \diff p \, D(p) =1$, characterizes the transverse coherence of the ensemble. We assume the initial ensemble to be incoherent on the scale of the grating period $d$, i.e.~the function $\widetilde{D}(s)$ is sharply peaked around $s=0$ and nonzero only for arguments $|s| \ll d$.
This is equivalent to saying that the momentum distribution $D(p)$ is broad and approximately constant on the scale of the grating momentum,
i.e. $\Delta p \gg \hbar / d$. This is the reason why a single grating does not suffice to observe quantum interference starting from such an initial state.

Once subjected to the first grating transformation, the initial state $w_0 (x,p) = D(p) / \Delta x$  undergoes a convolution in phase space $w_0 (x,p) \to \int \diff p_0 \, T^{(1)} (x,p-p_0) w_0 (x,p_0)$. The grating kernel $T^{(k)} (x,p)$ relates to the transmission function $t^{(k)} (x)$ of the $k$-th grating via
\begin{equation}
 T^{(k)} (x,p) = \frac{1}{2\pi \hbar} \int \diff s \, \exp \left( \frac{\rmi}{\hbar} ps \right) t^{(k)} \left( x - \frac{s}{2} \right) \left[ t^{(k)} \left( x + \frac{s}{2} \right) \right]^{*}. \label{eqn:T}
\end{equation}
After G$_{1}$ the state propagates freely during the time $T_1$. This corresponds to a shearing in phase-space, $x \to x - p T_1 / m$.
The same convolution and propagation transformation then applies to the subsequent diffraction due to G$_{2}$ and the following evolution during $T_2$. In order to obtain the final position distribution we integrate the Wigner function $w_{T_1 + T_2} (x,p)$ over
the momentum to get the spatial probability density
\begin{eqnarray}
 w_{T_1 + T_2} (x) &= \frac{1}{\Delta x} \int \diff p \diff p_1 \diff p_0 \, T^{(2)} \left( x - \frac{p T_2}{m}, p - p_1 \right) \nonumber \\
& \qquad \times T^{(1)} \left( x - \frac{p T_2 + p_1 T_1}{m}, p_1 - p_0 \right) D(p_0) . \label{eqn:w3T}
\end{eqnarray}
Due to the periodicity of the grating kernels (\ref{eqn:T}), $w_{T_1 + T_2} (x)$ is a $d$-periodic oscillatory function in $x$. It can be stated in terms of the Talbot-Lau coefficients of the $k$-th grating
\begin{equation}
 B_{n}^{(k)} (\xi) = \sum_{j=-\infty}^\infty b_j^{(k)} \left( b_{j-n}^{(k)} \right)^{*} \exp \left[ i \pi \xi \left( n - 2j \right) \right] \label{eqn:B},
\end{equation}
which involves the Fourier coefficients from Equation (\ref{eqn:bn}). We allow for different laser powers, i.e.~different parameters $n_0^{(k)}, \phi_0^{(k)}$, and following the procedure described in~\cite{Hornberger2009} we arrive first at an expression for the Talbot-Lau coefficients $B_{n}^{(k)} (\xi) $ (Eq.~(\ref{eqn:TLcoeffgeneral})) and then at the spatial probability density
\begin{eqnarray}
 w_{T_1 + T_2} (x) &= \frac{1}{\Delta x} \sum_{n,\ell=-\infty}^\infty \exp \left( \frac{2\pi \rmi \ell x}{d} \right) \widetilde{D} \left( \frac{d (n T_1 + \ell T_2)}{T_{\rm T}} \right) \nonumber \\
& \qquad \times B_{n}^{(1)} \left( \frac{n T_1 + \ell T_2}{T_{\rm T}} \right) B_{\ell - n}^{(2)} \left( \frac{\ell T_2}{T_{\rm T}} \right) . \label{eqn:w3}
\end{eqnarray}

\subsection{Resonance approximation}
A clear quantum interference pattern can be observed in the density distribution (\ref{eqn:w3}) when the pulses are at least separated on the order of the Talbot time. In this case the sharply peaked function $\widetilde{D}$ reduces the range of summation indices $(n, \ell)$ that contribute significantly to  $w_{T_1 + T_2} (x)$.
Only those $(n,\ell)$ count, which fulfill $|n T_1 + \ell T_2| \ll T_{\rm T}$. For $T_1, T_2 \simeq T_{\rm T}$ this generally implies a unique integer $n$ for each $\ell$~\cite{Nimmrichter2008b}. Here, we restrict ourselves to the case, where the delay $T_2$ after the second grating pulse is close to an integer multiple of the first pulse separation $T_1 = T$, $T_2 = N T + \tau$, with $|\tau| \ll T$. The double summation is then simplified by the resonance approximation
\begin{equation}
 \widetilde{D} \left( \frac{d (n T_1 + \ell T_2)}{T_{\rm T}} \right) \approx \delta_{n,-N\ell} \widetilde{D} \left( \frac{\ell d \tau}{T_{\rm T}} \right) .
\end{equation}
We then arrive at
\begin{equation}
 \fl w_{(N+1)T} (x) = \frac{1}{\Delta x} \sum_{\ell=-\infty}^\infty  \widetilde{D} \left( \frac{\ell d \tau}{T_{\rm T}} \right) B_{-N\ell}^{(1)} \left( \frac{\ell \tau}{T_{\rm T}} \right) B_{(N+1)\ell}^{(2)} \left( \frac{\ell (NT+\tau)}{T_{\rm T}} \right) \exp \left( \frac{2\pi \rmi \ell x}{d} \right) \label{eqn:w3res}
\end{equation}
for the position distribution immediately before the third laser pulse. In the resonant case, $\tau=0$, we find the ideal Talbot-Lau interference pattern as described by (\ref{eqn:w3TL}) for $N=1$.
As noted previously, the Talbot-Lau coefficients of the first grating $B_{n}^{(1)} (0)$ then reduce to the Fourier components (\ref{eqn:ATL}), and the first pulse serves as a classical mask without any phase modulation. This also shows that for an interference pattern to be formed, the first grating must not be a pure phase grating.

This is, however, no longer true for $\tau \neq 0$, where a transient near-field interference effect emerges because of the phase modulation at both the first and the second grating. This phenomenon was observed in Talbot-Lau interferometry of thermal atoms using pure phase gratings~\cite{Cahn1997,Turlapov2005}.
The maximum time span for such a transient interference signal is limited by the width of the initial momentum distribution, $|\tau| \lesssim m d / \Delta p$. Given a typical velocity spread of $\Delta v = \Delta p / m = 1\,$m/s, transient interference effects would occur for time shifts $\tau$ of mostly a few hundred nanoseconds with optical phase gratings.
Absorptive gratings, on the other hand, offer a resonant Talbot-Lau interference effect with high contrast, as shown in Section \ref{sec:numerics} and discussed below.

\subsection{Fringe shifts and the role of the third grating}
\label{sec:fringeshifts}
In our previous derivation of (\ref{eqn:w3res}) we assumed the particles to propagate freely between the grating pulses.
Now, we admit a constant force $F=ma$ along $x$ acting during the entire pulse sequence.
The propagation in the force field then causes an accelerated shearing of the Wigner function
\begin{equation}
 w_t (x,p) = w_0 \left( x - \frac{p t}{m} + \frac{a}{2} t^2 , p - m a t \right).
\end{equation}
Our earlier calculation can be redone for all pulses and reveals a shift $x \to x + \delta x$ of the density pattern (\ref{eqn:w3res}) at time $(N+1)T$, given by
\begin{equation}
 \delta x = - \frac{a}{2} N(N+1)T^2 . \label{eqn:deltax}
\end{equation}
Since all three grating laser beams are retro-reflected by the same fixed mirror, we may now exploit the fringe shift, Eq.~(\ref{eqn:deltax}), to scan the interference pattern by tuning the time separation. This will shift the fringe coordinate $x_S$ and modify the Fourier components of the detection signal after the third pulse,
\begin{equation}
 \fl S(x_S) = \int \diff x \, w_{(N+1)T} (x + x_S) \left|t^{(3)} (x) \right|^2 = \int \diff x \, w_{(N+1)T} (x) \left|t^{(3)} (x-x_S) \right|^2. \label{eqn:sigmask}
\end{equation}
In practice, the Earth's gravitational acceleration $a=g$ provides a highly homogeneous and constant force that can be easily used to shift the fringe pattern if the $x$-direction is chosen to have a vertical component. Alternatively, one may also use electrostatic fields.

When the signal is recorded as a function of $T$ it will no longer be strictly periodic. In practice, however, the time variation  $d / N (N+1) gT$ required to shift the pattern by about one grating period $d$ can be made small compared to the Talbot time $T_{\rm T}$ by increasing either the pulse separation time $T$ or the mass of the particles. The sinusoidal visibility then remains a good measure.
After inserting (\ref{eqn:w3res}) into (\ref{eqn:sigmask}), we obtain the final signal
\begin{equation}
 \fl S (x_S) = \sum_{\ell=-\infty}^\infty  \widetilde{D} \left( \frac{\ell d \tau}{T_{\rm T}} \right) B_{-N\ell}^{(1)} \left( \frac{\ell \tau}{T_{\rm T}} \right) B_{(N+1)\ell}^{(2)} \left( \frac{\ell (NT+\tau)}{T_{\rm T}} \right) B_{-\ell}^{(3)}(0) \exp \left( \frac{2\pi \rmi \ell x_S}{d} \right). \label{eqn:sigres}
\end{equation}
Its fringe contrast is
\begin{equation}
 \V_{\sin} = 2 \left| \frac{B_{-N}^{(1)} \left( \tau / T_{\rm T} \right) B_{(N+1)}^{(2)} \left( (N T+\tau) / T_{\rm T} \right) B_{-1}^{(3)}(0)}{B_{0}^{(1)}(0) B_{0}^{(2)}(0) B_{0}^{(3)}(0)} \, \widetilde{D} \left( \frac{ d \tau}{T_{\rm T}} \right) \right|. \label{eqn:vissinres}
\end{equation}
This shows that it is favorable to keep the pulse delays $T_1$ and $T_2$ equal, i.e.~to set $N=1$: Any larger $N$ increases the index of the Talbot-Lau coefficients (\ref{eqn:TLcoeffgeneral}) and thus reduces the maximum fringe contrast. In addition, the simulations of Section \ref{sec:numerics} show that a high visibility is realized with ionizing gratings in the resonant case $\tau=0$. Equation (\ref{eqn:vissinres}) suggests that this  deteriorates rapidly with increasing $|\tau|$ because of the sharply peaked Fourier transform $\widetilde{D} (s)$ of the momentum distribution.
We therefore consider the resonant case $\tau=0$ with equal pulse timing, $N=1$, in the following. 
The signal expression (\ref{eqn:sigres}) and the visibility (\ref{eqn:vissinres}) then reduce to Equations (\ref{eqn:sigTL}) and (\ref{eqn:vissinTL2}), respectively.

In an alternative `inverse configuration' of G$_{3}$, where the ions are counted instead of the neutral clusters, we have to replace $|t^{(3)} (x)|^2$ by $1 - |t^{(3)} (x)|^2$. The grating coefficients in (\ref{eqn:sigres}) then change from Eq.~(\ref{eqn:ATL}) into $B_n^{(3)}(0) = \delta_{n,0} - \exp (-n_0^{(3)}/2) I_n \left( - n_0^{(3)}/2 \right)$.

\section{Challenges and limitations}

\subsection{Rayleigh scattering in the grating} \label{sec:scattering}
So far the influence of the optical gratings was treated as a coherent transformation that modulates the cluster wave function through the complex transmission function (\ref{eqn:t}). However, in addition to the absorption of photons and their virtual scattering within the laser wave the diffracted nanoparticles may also give rise to Rayleigh scattering, where the photon is re-emitted elastically. This process causes momentum diffusion \cite{Chapman1995,Kokorowski2001} and thus deteriorates the interference pattern. 

For short laser pulse durations we can treat the Rayleigh scattering independently of the coherent grating transformations. We model it by a Lindblad-type master equation for the density matrix~\cite{Kazantsev1990,Dyrting1994a,Domokos2001}
\begin{equation}
 \fl \partial_t \rho = \gamma_{\rm R} \left[ \int_{|\vu|=1}
\!\!\!\!\!\!\!\!\diff \vu \, N(\vu)
\cos \left( k_{\rm L} \ox \right) e^{-\rmi k_{\rm L} u_x \ox} \rho e^{\rmi k_{\rm L} u_x \ox} \cos \left( k_{\rm L} \ox \right) - \frac{1}{2} \left\{ \rho, \cos^2 \left( k_{\rm L} \ox \right) \right\} \right] \label{eqn:scattME}.
\end{equation}
The position operator is denoted by $\ox$.
The equation describes the elastic scattering of a single photon from the diffracting laser field into a random direction $\vu$.
The corresponding momentum recoil is represented by the operators for the standing-wave mode $\cos k_{\rm L} \ox$ and the plane wave modes $\exp \left( \rmi k_{\rm L} u_x \ox \right)$. The number of recoils is determined by the total scattering rate $\gamma_{\rm R}$ at the antinodes of the light field, and $N(\vu)$ gives the distribution of scattering directions, as described by the dipole pattern $N(\vu) = 3 \sin^2 \vartheta / 8 \pi$. Here $\vartheta$ is the angle between $\vu$ and the polarization vector of the laser field.

We can solve the master equation (\ref{eqn:scattME}) in the position representation. The density matrix then transforms according to $\la x |\rho|x' \ra \to \eta (x,x') \la x |\rho|x' \ra$, with the decoherence function
\begin{equation}
 \fl \eta (x,x') = \exp \left[ \frac{n_{\rm R}}{2} \left( 2\int_{|\vu|=1}
\!\!\!\!\!\!\!\!\diff \vu \, N(\vu) e^{ik_{\rm L} u_x (x'-x)} \cos k_{\rm L} x \cos k_{\rm L} x' - \cos^2 k_{\rm L} x - \cos^2 k_{\rm L} x' \right) \right]. \label{eqn:eta}
\end{equation}
It is trace-preserving, $\eta(x,x)=1$, and it only acts on the spatial coherences. The mean number of scattered photons at the antinodes of the $k$th grating is related to the number of absorbed photons via the ratio of the associated cross sections, $n_{\rm R}^{(k)} = n_0^{(k)} \sigma_{\rm R} / \sigma_{\rm abs}$. In phase space representation, decoherence due to elastic scattering (\ref{eqn:eta}) is then represented by the integral kernel
\begin{eqnarray}
 T_{\rm R}^{(k)} (x,p) &= \frac{1}{2\pi \hbar} \int \diff s \, \exp \left[ \frac{\rmi}{\hbar} ps + n_{\rm R}^{(k)} \eta \left( x - \frac{s}{2}, x + \frac{s}{2} \right) \right] \nonumber \\
 &= \sum_{n=-\infty}^\infty \exp \left( \frac{2\pi \rmi n x}{d} \right) \int \diff s \, \exp \left( \frac{\rmi}{\hbar} ps \right) R_n^{(k)} \left( \frac{s}{d} \right) \label{eqn:Ts},
\end{eqnarray}
which has to be applied at each grating pulse $k$ in addition to the coherent grating kernel (\ref{eqn:T}). Again, the periodicity of the kernel in $x$ allows us to expand it as a Fourier series, where the closed expression for the Fourier coefficients
\begin{eqnarray}
R_n^{(k)} (\xi) &= \exp \left\{ \frac{ 3 n_{\rm R}^{(k)}}{4} \left[ \cos \pi \xi \left( j_0 \left( \pi \xi \right) - \frac{j_1 \left( \pi \xi \right)}{\pi \xi} \right)  - \frac{2}{3} \right]\right\}  \nonumber \\
&\quad \times I_n \left[ \frac{3 n_{\rm R}^{(k)}}{4} \left( j_0 \left( \pi \xi \right) - \frac{j_1 \left( \pi \xi \right)}{\pi \xi} - \frac{2}{3} \cos \pi \xi \right)\right]
\end{eqnarray}
follows from a lengthy calculation involving combined integrals over trigonometric expressions and Bessel functions. The symbol $j$ denotes the spherical Bessel function of the first kind. Combining the coherent kernel (\ref{eqn:T}) and the scattering kernel (\ref{eqn:Ts}) results in the modified Talbot-Lau coefficients
\begin{equation}
 \hat{B}_n^{(k)} (\xi) = \sum_{j=-\infty}^{\infty} R_{n-j}^{(k)} (\xi) B_j^{(k)} (\xi) \label{eqn:Bs}
\end{equation}
for each grating. Since $\hat{B}_n^{(k)} (0) = B_n^{(k)} (0)$ the Rayleigh scattering acts as a purely decohering agent and does not influence the periodic masking  by the gratings. This means, in particular, that only the treatment of the second grating needs to be modified.

\begin{figure}
\includegraphics[width=12cm]{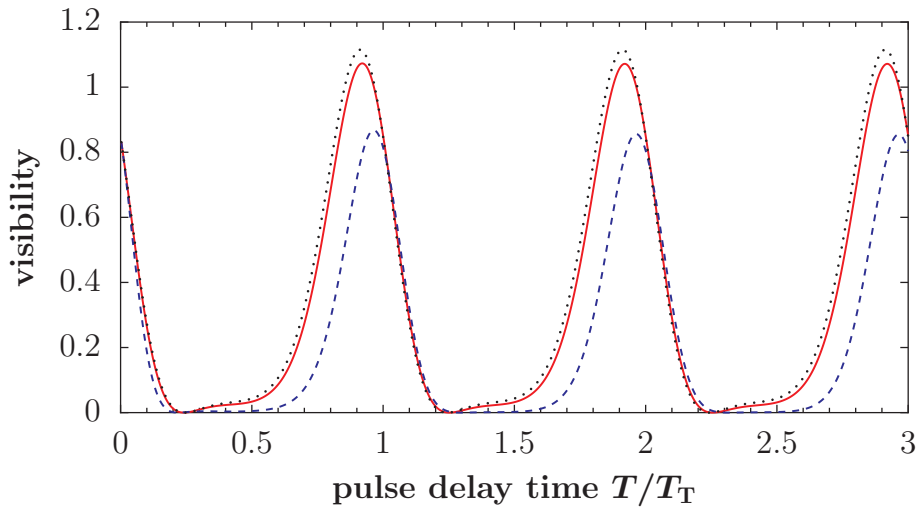}
\caption{\label{fig:scatteringmass} Calculated fringe visibility including Rayleigh scattering of laser photons as function of $T/T_{\rm T}$ for a fixed $n_0^{(2)}$ and gold clusters of different masses. The solid and the dashed lines correspond to $m=10^8\,$amu and $m=10^9\,$amu, respectively. The dotted line reproduces the quantum contrast curve of Fig.~\ref{fig:visqmklass}(a) without Rayleigh scattering, which is a good approximation for cluster masses $\lesssim 10^8\,$amu. }
\end{figure}

With growing cluster size the role of elastic light scattering becomes increasingly important. Since the scattered photons have the same short wavelength as the grating light, they can decohere the previously delocalized matter waves. We may disregard this effect only as long as the number of scattering events $n_{\rm R}^{(2)}$ is small compared to the number of absorption processes $n_0^{(2)}$. Their ratio is given by their total cross sections
\begin{equation}
\frac{n_{\rm R}^{(2)}}{n_0^{(2)}} = \frac{\sigma_{\rm R}}{\sigma_{\rm abs}} = \frac{2}{9} \frac{ (\varepsilon_1 -1)^2 + \varepsilon_2^2 }{\varepsilon_2} \left( k_{\rm L} R \right)^3 = \frac{4 \pi^2}{3} \frac{ (\varepsilon_1 -1)^2 + \varepsilon_2^2 }{\varepsilon_2} \frac{m}{\rho \lambda_{\rm L}^3} 
\end{equation}
and it is proportional to the mass, assuming that the dielectric function $\varepsilon = \varepsilon_1 + \rmi \varepsilon_2$ and the density $\varrho$ are constant. The influence of Rayleigh scattering on the interference contrast is shown in Fig.~\ref{fig:scatteringmass}, where the expected sinusoidal fringe visibility is plotted as a function of the pulse delay around the Talbot time, both for gold clusters of $m=10^8\,$amu (solid line) and for $m=10^9\,$amu (dashed line).
The power of the UV laser gratings is adjusted such that the absorption parameters are kept constant at $n_0^{(1)}=n_0^{(2)}=n_0^{(3)}=8$  (as in Section \ref{sec:numerics}).
While the solid line still resembles the predicted visibility without Rayleigh scattering (dotted line), the overall contrast of the dashed curve is somewhat suppressed due to the increased scattering cross-section of $\sigma_{\rm R} = 0.9 \sigma_{\rm abs}$ at $m=10^9\,$amu.  This corresponds to a gold cluster of 55\,nm diameter, a size where the point particle approximation ceases to be valid. It also exceeds by two to three orders of magnitude the mass limit where the experiments have to be modified in order to compensate for the influence of gravity (see Sec.~\ref{sec:averaging}). 

\subsection{Collisional decoherence and thermal emission}

Another concern in interferometry with massive particles is their uncontrolled interaction with the environmental degrees of freedom. The most prominent causes of decoherence are collisions with residual gas particles and the emission or scattering of thermal radiation~\cite{Hornberger2003,Hackermueller2004}.
We find that the unperturbed interference of a $10^6\,$amu gold cluster after a delocalization time of $2T_{\rm T} \simeq 30\,$ms requires that the background gas pressure is kept at $p \lesssim 10^{-9}\,$mbar and the internal cluster temperature at $T \lesssim 1000\,$K. Both values can be readily achieved in an experiment.
For this estimate we chose N$_2$ as the background gas and use the London dispersion formula to compute its interaction strength with the gold clusters, assuming 
the ionization potential of N$_2$ to be 15.6\,eV and its static polarizability $\alpha_{N_2}=4\pi\varepsilon_0\times 1.74 \AA^3$.
The dielectric function of gold at thermal photon wavelengths is here assumed to follow Drude's model with a work function of $W=5.4$\,eV, a plasmon frequency of $\omega_P=1.3\times 10^{16}$\,Hz and a resonance width of $\Gamma_P=1.1 \times 10^{14}$\,Hz. The static polarizability of gold is taken to be that of an ideal metal cluster.

\subsection{Fringe averaging on Earth}
\label{sec:averaging}

As seen in Sec.~\ref{sec:fringeshifts}, gravity can cause a fringe shift if the direction of the force is parallel to the diffraction grating vector.  For usual interferometers this phase shift is dispersive, i.e.~larger for clusters of lower velocity and longer falling time. Even a small width of the cluster velocity distribution
can then cause a rapid fringe averaging in the plane of G$_3$. 
This problem can be completely eliminated in an OTIMA interferometer where all clusters are exposed to the gravitational field for exactly the same amount of time. All phase shifts are therefore equal and one can still expect high contrast interference, independent of the grating orientation in space.

What remains is the issue of the classical falling distance. At $10^6$\,amu a cluster needs a passage time of about 30\,ms, assuming F$_2$ laser gratings with a period of $78.5\,$nm. During this time it will fall by about 4.6\,mm. At $m=10^7$\,amu and a time of 300\,ms the free fall distance reaches already half a meter. This illustrates that the force of gravity will have to be compensated at some point by electrical or magnetic fields or to be eliminated in a microgravity environment.

In contrast to gravity, the Coriolis force $F_{\rm C}=2m \vec{v}\times\vec{\Omega}$ depends on the cluster velocity both in modulus and direction. It cannot be compensated by clever timing.  However, OTIMA offers still an advantage over a fixed-length interferometer because it allows us to align the mirror orientation normal to $\vec{\Omega}$. The resulting Coriolis force is then intrinsically orthogonal to the grating vector $\vec{d}$, such that the acceleration does not contribute to the fringe shift.

With growing masses and Talbot times the interferometer becomes more sensitive to vibrations. For low-frequency vibrations, the total fringe shift depends on the position change $\Delta d$ of  each individual grating according to  $\Delta d = \Delta d^{(1)} - 2\Delta d^{(2)} + \Delta d^{(3)} $.
The single mirror design of the proposed OTIMA interferometer (Figure~\ref{fig:pulsescheme}) ensures that many conceivable contributions drop out, such as quasi-static mirror shifts and mirror tilts. All dynamical mechanical effects have to be suppressed on the level of about 5-10\,nm, which can be done using established spring suspension systems.

\section{Experimental considerations} \label{sec:exp}
Having described the general idea, we now discuss the compatibility of existing beam sources, diffraction elements and detectors with our suggested scheme and we start with the availability of single-photon ionization (SPI) gratings~\cite{Reiger2006} for various materials.

\subsection{Grating requirements}
For our present proposal we consider only particles, for which efficient single-photon ionization (SPI) has already been observed~\cite{Heer1993,Wahl1994,Koch2007}, such as metal clusters. The cluster ionization energy usually exceeds the work function of the bulk by less than 1-2\,eV. It decreases with the particle radius like $E_{\mathrm{ion}} \simeq W+ 0.42 e^2/4\pi\varepsilon_0 R$ and approaches the bulk value at 20 to 100 atoms per cluster~\cite{Heer1993}. Dozens of materials are therefore suitable candidates for SPI diffraction and detection, provided that the photon energy is higher than 5\,eV, i.e.~$\lambda\le 250$\,nm.

Although SPI yields are not quantitatively known for all materials, the process has been shown to be dominant in several systems, also in competition with fragmentation and photoemission. This applies to semiconductor nanoparticles~\cite{Schaefer1997} as well as to metal clusters~\cite{Wucher1996}.
In our simulations (Figs.~\ref{fig:visqmklass},~\ref{fig:visbeta},~\ref{fig:grating13}) we have assumed a yield  of $\eta_{\rm SPI} \simeq 1$. 

In the following we distinguish interferometry with particles below 20,000\,amu and from $10^4$ to $10^6$\,amu, since they may require different implementations of the same idea. 

We envisage a de Broglie wavelength of $\lambda_{dB} \simeq 5\times 10^{-13}$\,m, to fit the entire interferometer on the same two-inch mirror. This is compatible with a velocity of 40\,m/s for clusters with $m\le 20,000$\,amu. This layout is driven by the desire to have a short and ultra-stable base for all laser gratings,  which is important for high signals and with regard to a high common-mode rejection for phase fluctuations related to vibrations and rotations of the mirror mount.

For this setting, a F$_2$ excimer laser is  well adapted. It emits a few hundred VUV light pulses per second with an energy of more than 1\,mJ and a duration of less than $5\,$ns.
Retroreflection of the laser light generates a diffraction grating with a period of $d=78.5$\,nm.
The F$_2$ laser wavelength is fixed at $\lambda=157.63094(10)$\,nm with a line width of less than 1\,pm~\cite{Sansonetti2001}.  The gain profile of fluorine allows, in principle, a five times weaker line at $\lambda=157.52433(10)$\,nm, as well. The \emph{longitudinal} coherence length of more than 1\,cm, guarantees a sufficient grating periodicity up to a distance of 2\,mm from the retro-reflecting mirror. This covers the entire width of the cluster beam and about 10$^5$ grating periods.

The \emph{transverse coherence}  of excimer lasers is limited and often determined by  diffraction of the light beam at the laser outcoupler window. Following the van Cittert Zernike theorem~\cite{Born1993} and also recent measurements  the transverse coherence is estimated to be 80\,$\mu$m in a meter behind the laser window~\cite{Dyer2009}.
The homogeneity of the standing light wave is dominated by the quality of the mirror surface.
A flatness of better than $d/10$ is available for VUV wavelengths.
Given this, even the limited transverse coherence is still sufficient for the formation of a homogeneous diffraction grating, provided that the laser beam divergence is limited to about 1\,mrad.

All this suggests that a F$_2$ laser is the ideal basis for building a compact and rugged OTIMA interferometer which allows one to address a large range of materials. This flexibility is bought, however, at the price of working with vacuum ultraviolet laser light, which requires purged beam lines and specialized optics.

The short laser pulse width is an advantage for  precision measurements but also entails some further consequences related to the fact that the standing light wave needs some time to form and that it also experiences an overall temporal amplitude modulation.
The finite speed of light is the reason why the clusters will interact with a running wave during a fraction of the laser pulse duration. For a cluster beam confined to within 1\,mm from the mirror surface, this running wave is limited to 0.5\,\% of the total pulse energy. Also the low reflectivity of even the best mirrors which are available for 157\,nm ($R\ge98$\,\%)
causes the cluster beam to be exposed to a running light wave of 2\,\% the total pulse energy.
This slightly affects the final count rate but not the visibility, since only the neutral particles reach the detector.

Amplitude variations of the laser light field, realistically on the order of 3 to 5\,\%, modulate the population of different diffraction orders but they cannot destroy the interference pattern, since  the fringe positions are solely determined by the laser wavelength.
Also the pointing instability of the grating lasers is negligible if one can ensure that all detected particles have interacted with all three gratings. A critical point is the accurate timing, i.e a low jitter, of the laser gratings. With existing technologies a timing accuracy of 1\,ns can be achieved, which is sufficient for the purpose.

Although the Talbot time $T_T=d^2m/h$ is intrinsically independent of the particle's velocity, clusters of different speed will travel different distances between the gratings of the same interferometer. Therefore, the cluster beam must be sliced into packets (Figure \ref{fig:pulsescheme})  such that all particles of a batch always interact with the same spatially extended laser pulse, independent of their velocity. The combination of a chopped source with a pulsed TOF-MS detector will, in practice, select a  velocity band of $\Delta v / v \simeq 2\%$ in the proposed experiments up to $10^4$\,amu. A selection of de Broglie wavelengths is then achieved in combination with the detecting mass spectrometer.  In principle, an interferogram may then even be recorded without scanning either velocity, time or grating position, provided the source emits a sufficiently dense set of cluster masses.

In the range of $10^6$\,amu the particle cloud will already expand quite significantly during a total interferometer time of 30 to 60\,ms, even if we assume a cluster velocity of 1\,m/s, which is a challenge in itself. We thus require a laser with a spectral coherence of $\Delta \nu \le 0.1$ cm$^{-1}$, as for instance provided by a grating stabilized ArF laser at 193 nm, the fifth harmonic of a seeded Nd:YAG laser at 213\,nm, a frequency doubled dye lasers or an optical parametric oscillator in the wavelength range between 210 and 250\,nm.
Since the absorption cross sections of metal clusters grow linearly with the mass for a given atomic species, 
their higher values will then compensate for the intensity loss when the UV laser beam must be expanded to cover the growing cluster cloud.
Finally, in the high-mass range the temporal envelope of the laser beams grows in relevance. The expanded cloud will now see a running wave for up to several percent of the total pulse width.
This time may be reduced by making a standing light wave from counter propagating laser beams.
 
\subsection{Source properties}

On the source side, different options are at hand for different mass ranges:
Interestingly, even an {\em effusive beam source} may match the requirements of an OTIMA  interferometer.  The most probable velocity for $m=10^6$\,amu at $T$=500\,K is  $v_{\mathrm{mp}} \simeq  3\,$m/s and recent experiments have shown that there is still hope for the thermal volatilization of highly massive but chemically tailored organic molecules~\cite{Gerlich2011}.

However, for the purpose of our present proposal we restrict ourselves to a {\em cluster condensation source}, which first volatilizes the atoms and then recondenses them in a cold noble gas stream. The initial atomization may be done thermally, for some materials such as alkali atoms~\cite{Heer1993} and fullerenes~\cite{Martin1984}, or using a magnetron sputter technique for a large range of metal clusters with higher melting points~\cite{Haberland1991}. A gas aggregation source ejects a broad distribution of clusters, ranging from a few up to several ten thousand atoms per cluster with a brilliance of up to $10$\,mg~srad/s. 

Own preliminary studies show that under conditions comparable to OTIMA we can count about 100 cluster ions on the TOF-MS MCP in a time window of $200\,\mu$s. To achieve an adequate visibility each laser grating has to reduce the cluster beam by a factor of three. After the interferometer we thus retain 4\% 
of the initial beam, i.e.~$600$ ions per second with virtually no background during the measurement period, which is sufficient for the purpose of interferometry.

In the setup of Figure~\ref{fig:pulsescheme} the cluster velocity is determined by the flow of the cold carrier gas, which is usually a mixture of Helium and Argon. The most probable cluster velocity in a nitrogen cooled setup amounts to about  $v_{\mathrm{mp}}=300$\,m/s. This can be lowered by a cryogenic cooling stage, using a neon buffer gas at 30\,K ~\cite{Patterson2009} to achieve a cluster velocity in the range of less than $100$\,m/s, if we include the relative velocity slip of massive clusters in light noble gases~\cite{Rosemeyer2001}. From this distribution one may still select the low-velocity tail to get to particles in the range of 40 to 50\,m/s. Moreover, experiments aimed at surface cleaning also showed that laser acoustic desorption is capable of generating nanoparticle beams with speeds in the range of 10 to 50\,m/s~\cite{Geldhauser2007}.

Many complementary methods are still conceivable. Of particular interest are those techniques that focus on the improvement of cooling for freely suspended mesoscopic particles~\cite{Chang2010,Romero2010,Romero2011,Nimmrichter2010}.  Although most published proposals are still best adapted to small source volumes and a few particles, a rapidly growing community is currently investigating methods for preparing mesoscopic isolated particles, which may eventually also be coupled to our proposed interferometer scheme.
All this suggests that there are ample possibilities for further progress to masses up to $10^6$\,amu.

\section{Conclusions and Outlook} \label{sec:outlook}

The proposed OTIMA-interferometer is universal in the sense that a single experimental setup will enable experiments with a large class of nanoparticles, ranging from single atoms to metal clusters, semiconductor nanocrystals, and possibly even some biomolecular complexes, as long as ionization or neutralization is the dominant response to the absorption of a single photon. This covers a mass range from 6 and 7\,amu for lithium isotopes to $10^6$\,amu for cold giant clusters.

The setup is promising for testing the foundations of wave mechanics, and its intrinsic features, such as high common mode rejection, non-dispersiveness and good control over the grating pulse times, are well-adapted to interference-assisted cluster metrology. The OTIMA concept could thus enable new insights into cluster physics, including collisional and absolute photoabsorption cross sections, electric and magnetic properties etc.

Our calculations show that a high interference contrast may be expected for a wide range of masses under otherwise very similar conditions. This could become relevant in explorations of the equivalence principle as well as in experiments that probe currently speculative fundamental limitations of the Schr{\"o}dinger equation~\cite{Bassi2003, Diosi2004, Penrose1996,Wang2006,Carlip2008,Adler2009}.

In contrast to many other experiments where source purity is a value in itself, the OTIMA interferometer can profit from the broad distribution of particles ejected by a magnetron clusters source~\cite{Haberland1991}.
Performing experiments with different cluster types under otherwise identical conditions provides a high degree of `common mode noise rejection' and a good way of identifying systematic phase shifts under variations of mass, baryon composition, geometrical structure or other cluster properties.

\ack
We thank the Austrian Science Fund (FWF) for support within
the projects Wittgenstein Z149-N16 and  DK-W1210 CoQuS.
We also acknowledge financial support within the \emph{MIME} project of the ESF EuroQUASAR program, through FWF and Deutsche Forschungsgemeinschaft (DFG).

\section*{References}

%\bibliographystyle{iopart-num}
%\bibliography{mqo,addlit}

%\providecommand{\newblock}{}

\end{document}